\documentclass[
  twocolumn,
  prl,
  showpacs,
  amsmath,
  amssymb,
  superscriptaddress,
  floatfix
]{revtex4}

\usepackage{bm}
\usepackage{dsfont}
\usepackage{graphicx}
\usepackage{pifont}
\usepackage{textcomp}
\usepackage{gensymb} 
\usepackage{hyperref}

\usepackage{color}

\newcommand{\be}{\begin{equation}}
\newcommand{\e}{\end{equation}}
\newcommand{\beml}{\begin{subequations}}
\newcommand{\eml}{\end{subequations}}
\newcommand{\beq}{\begin{eqnarray}}
\newcommand{\eq}{\end{eqnarray}}
\newcommand{\ba}{\begin{array}}
\newcommand{\ea}{\end{array}}
\newcommand{\bpm}{\begin{pmatrix}}
\newcommand{\epm}{\end{pmatrix}}
\newcommand{\bc}{\begin{cases}}
\newcommand{\ec}{\end{cases}}
\newcommand{\lt}{\left}
\newcommand{\rt}{\right}

\newcommand{\la}{\langle}
\newcommand{\ra}{\rangle}
\newcommand{\ep}{\varepsilon}

\newcommand{\bb}{\boldsymbol}

\DeclareMathOperator{\sign}{sign}

\begin{document}
\title{Metal-Insulator Transition in Graphene on Boron Nitride}

\author{M.~Titov}
\affiliation{
Radboud University Nijmegen, Institute for Molecules and Materials, NL-6525 AJ Nijmegen, The Netherlands}

\author{M.~I.~Katsnelson}
\affiliation{
Radboud University Nijmegen, Institute for Molecules and Materials, NL-6525 AJ Nijmegen, The Netherlands}

\begin{abstract}
Electrons in graphene aligned with hexagonal boron nitride are modelled by Dirac fermions in a correlated random-mass landscape subject to a scalar- and vector-potential disorder. We find that the system is insulating in the commensurate phase since the average mass deviates from zero. At the transition the mean mass is vanishing and electronic conduction in a finite sample can be described by a critical percolation along zero-mass lines. In this case graphene at the Dirac point is in a critical state with the conductivity $\sqrt{3}e^2/h$.  In the incommensurate phase the system behaves as a symplectic metal.  
\end{abstract}


\maketitle

%%%%%%%%%%%%%%%%%   Introduction %%%%%%%%%%%%%%%%%%%%%%%%%%%%%%%%%%

Several years of intense development singled out the hexagonal form of boron nitride (hBN) as a unique insulating substrate for graphene \cite{Dean10} (for a review, see Refs.~\onlinecite{Geim13,Yank14}). High values of electron mobility in graphene on hBN \cite{Dean10}, exceeding those in suspended samples at room temperature \cite{Castro2010}, are now routinely achievable in a lab, paving the way to a variety of high-quality graphene devices \cite{Geim13,Yank14,Britnell12}. A number of interesting discoveries, including, e.g., giant magnetodrag \cite{Titov13} and non-local transport \cite{Abanin11}, have been made using graphene/hBN samples. The mutual orientation of graphene and hBN lattices in these high-mobility samples has been entirely random. 

More recently it became technologically possible to control the orientation of the graphene lattice on the hBN substrate \cite{Yankowitz12}. This development has already led to the discovery of ``Hofstadter butterfly'' physics \cite{ponomarenkoN13,deanN13,huntS13} and to the observation of insulating behaviour in some highly oriented samples \cite{Woods14}. 

It is understood both theoretically and experimentally that graphene and hBN lattices do not fully match on the atomic level even if their lattice orientations coincide.  Instead, a 1.8\% difference in lattice spacing between graphene and hBN makes it energetically favourable to develop local lattice distortions, which are seen as a moir\'e pattern with a period of 14\,nm \cite{Woods14,Sachs11}. The periodic lattice distortions with smaller periods are also observed in samples with a tiny orientation angle $\phi \lesssim 1\degree$ between the graphene and hBN lattice \cite{Woods14}. These samples are in the commensurate phase. No essential lattice reconstruction occurs in the incommensurate phase for larger orientation angles. 

In this Letter we focus on the metal-insulator phase transition in graphene at charge neutrality, which can accompany the incommensurate-commensurate phase transition in sufficiently clean samples. Insulating behaviour has been so far observed for a few samples doped to charge neutrality in the commensurate phase \cite{Woods14}. We interpret the observation as the result of the mean band gap opening in graphene (mean mass), which is induced by the proximity to the hBN. We also argue that the metal-insulator crossover can be modelled in a finite system by classical percolation of Dirac fermions in a long-range correlated random mass landscape.

As far as the electronic properties are concerned we may use the standard tight-binding model for graphene, assuming that the major effect of proximity to the hBN is described by external potentials $V_A(\bb{r})$ and $V_B(\bb{r})$ induced on graphene sub-lattices $A$ and $B$.  We further assume that lattice distortions are smooth on atomic scales. It is, then, convenient to define $V_0=(V_A+V_B)/2$ and $m=(V_A-V_B)/2$, and describe charge transport in graphene at zero temperature by the effective Dirac Hamiltonian
\be
\label{model}
H=\hbar v\, \bb{\sigma}\cdot \big(\bb{p}+\tau_z \bb{A}(\bb{r})\big) +\sigma_z\tau_z m(\bb{r})+V(r),
\e
where $v\approx 10^6$\,m/s is the Fermi velocity, the Pauli matrices $\bb{\sigma}=(\sigma_x,\sigma_y)$ and $\sigma_z$ act in the sub-lattice space, while the Pauli matrix $\tau_z$ acts in the valley space (for a general introduction to the Dirac fermion physics in graphene, see Refs.~\onlinecite{graphene-review,Katsnelson-book}). 

The strain induced in graphene is responsible for the appearance of the vector-potential-like term $\bb{A}$, and for the contribution to the scalar potential $V_s$. The long-range Coulomb impurities give another random contribution to the scalar potential $V_\textrm{imp}$, so that $V=V_0+V_s+V_\textrm{imp}$. Without loss of generality we assume that $\la V\ra=0$, where the brackets stay for the average over disorder. We study the model (\ref{model}) at zero chemical potential which corresponds to charge neutrality.  The inter-valley scattering, which can be induced by point-like impurities or other atomic defects, is completely ignored in the model (\ref{model}), since we restrict our attention to high mobility samples. 

A number of \textit{ab initio} studies have been undertaken recently to characterise the commensurate phase \cite{Sachs11,Gio07,Sakai11,Bokdam13,Merel14}. It has been established that the energetically favourable local configuration corresponds to the placement of boron against carbon and nitrogen against the void in the graphene lattice. Such a configuration would be entirely stable on large scales if the graphene and hBN lattices had an identical lattice spacing. In reality, the graphene lattice tends to be stretched to become commensurable with that of hBN, but the energy gained by bonding commensurability is insufficient to fully offset the energy costs of stretching. The interplay of the two mechanisms forms a complex lattice distortion, which consists of the starched honeycomb areas of the graphene lattice in the most favourable configuration (such that nitrogen is opposite to the void), which are separated by the areas with the second optimal configuration (nitrogen against carbon and boron against the void). This picture is indeed supported by local probe microscopy \cite{Woods14}. Recent atomistic simulations \cite{Merel14} show that the experimentally observable distribution of atomic displacements can only be explained by assuming that the C-N interaction is at least 2 times stronger than the C-B one. 

Thus, in the commensurate phase with a perfect lattice alignment, $\phi=0\degree$, a positive mass potential is induced inside moir\'e hexagons due to the interaction with the boron atom (the absolute choice of mass sign is irrelevant), while on the boundary between the moir\'e hexagons the mass is largely negative due to the interaction with the nitrogen atom. The average value of the band gap, i.e., the mean mass potential, can be measured, for instance, by the magneto-optic response. Such measurements have indeed confirmed the existence of a finite average mass in the commensurate phase \cite{Chen14}. 

Theoretically, the mean square deviation $\Delta=\sqrt{\la \delta m^2\ra }$ in the commensurate phase has been predicted to vary from $\Delta\sim$ 10\,meV to $\Delta \sim$ 300\,meV \cite{Sachs11,Gio07,Sakai11,Bokdam13} and a finite value of $\la m\ra$ in the commensurate phase has been found in Ref.~\cite{Bokdam13}. The enhancement of $\Delta$ due to electron-electron interactions has been also discussed qualitatively in Ref.~\onlinecite{Song13}. We explain below that the appearance of a finite mean mass is the main reason for the observed insulating behaviour in the commensurate phase. 

With the increase of orientation angle $\phi$ the regular pattern for the mass potential breaks down so that the mass can be modelled by a long-range correlated disorder potential with the correlation length $\xi$, the mean value $\la m\ra$, and the mean square deviation $\Delta$.  At some value of $\phi$ of the order of $1\degree$, the value $\la m\ra$ is close to zero, while the amplitude $\Delta$ is still of the order of $10$--$50$\,meV. Deep in the incommensurate phase, i.e. for $\phi\gg 1\degree$, the mass potential becomes completely negligible.

%%%%%%%%%%%%%%%%%%%%%%%%%%%%
%%%% fig:phase
%%%%%%%%%%%%%%%%%%%%%%%%%%%%
\begin{figure}[ht]
\centerline{\includegraphics[width=\columnwidth]{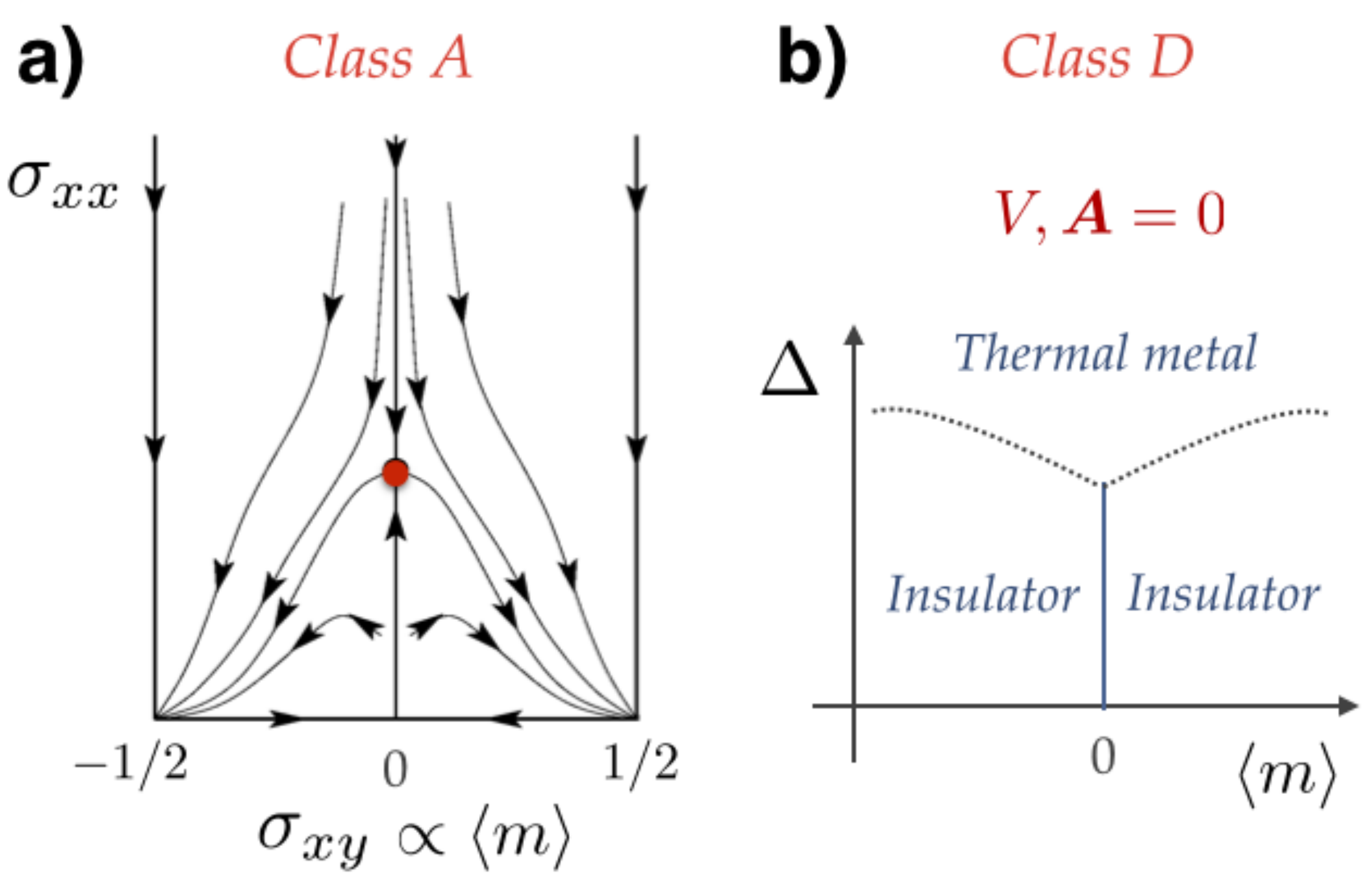}}
\caption{a) Flow diagram for the model (\ref{model}) with random $V$, $\bb{A}$, and $\bb{m}$ (class $A$) \cite{Pruisken85}. Conductivity flows to quantum Hall critical point $\sigma^\textrm{QHE}_{xx}\approx 0.57\times 4e^2/h$ provided $\la m\ra=0$ and to zero otherwise \cite{ostrovsky07,evers08,Gattenloehner14}. (b) Phase diagram in the class $D$ ($V,\bb{A} =0$). If $\Delta$ is smaller than a critical value, the conductivity flows to $\sigma^{D}_{xx}=4e^2/\pi h$ for $\la m\ra =0$ and to zero otherwise \cite{ostrovsky06,evers08}.}
%\vspace*{-0.5cm}
\label{fig:phase}
\end{figure}
%%%%%%%%%%%%%%%%%%%%%%%%%%%%% 

The decisive role of the average mass $\la m\ra$ for conduction is well documented for the model (\ref{model}) with a short-range correlated random mass potential \cite{Ludwig94,ostrovsky06,ostrovsky07,evers08,Bardarson10,medvedeva}. Indeed, the model with fully random $V$, $\bb{A}$, and $m$ belongs to the quantum Hall symmetry class $A$ \cite{Zirnbauer}. At the Dirac point such a system is known to have two insulating phases, $\la m\ra >0$ and  $\la m\ra<0$, separated by the quantum-Hall critical state at $\la m\ra=0$ as illustrated in Fig.~\ref{fig:phase}(a). Zero-temperature conductivity in the critical state flows to $\sigma^\textrm{QHE}_{xx} \sim 0.57 \times 4e^2/h$ in the thermodynamic limit \cite{Pruisken85,ostrovsky07,evers08,Gattenloehner14}. 

If the amplitudes of $V$ and $\bb{A}$ are negligible, the system belongs to the class $D$, which is characterised by the phase diagram depicted in Fig.~\ref{fig:phase}(b). (The existence of the thermal-metal phase in the model is, however, debated \cite{Bardarson10}.) The conductivity in this situation shows essentially the same behaviour as in the class $A$, except that its value at criticality, $\la m\ra =0$, is different: the conductivity flows to $\sigma^D_{xx}=4e^2/\pi h$ instead of $\sigma^\textrm{QHE}_{xx}$ \cite{ostrovsky06,evers08}. 

The correlations of mass potential cannot change the behaviour of the system in the limit of infinite system size. Thus, the model (\ref{model}) with any fluctuating $V$, $\bb{A}$, and $m$ is subject to Anderson localisation and gives rise to the insulating phase at charge neutrality unless $\la m\ra=0$.  Despite the mathematical rigour of the statement it might be impossible to observe the expected insulating behaviour experimentally due to a finite system size or finite temperature, i.e., in the situation that the localisation length becomes larger than either the system size or dephasing length.  

The high-quality graphene/hBN samples with controlled lattice alignment are restricted in size to a few microns. It is very likely, therefore, that  
the observed insulating behaviour originates in classical percolation. The existence of the percolation regime requires that the fluctuations of the scalar potential are suppressed, $\sqrt{\la V^2\ra} \ll \Delta$, while the correlation length of the mass potential is sufficiently long, $\xi \Delta/\hbar v \gg 1$.  The conditions define the percolating network of topologically protected chiral channels, which are given by the lines of zero mass.  

The origin of the chiral channels can be understood from the symmetry analysis. In the absence of magnetic field, the Hamiltonian (\ref{model}) yields the physical time-reversal symmetry: $\tau_y\sigma_y H^T \tau_y\sigma_y =H$.  Since the Hamiltonian is block diagonal we can, however, consider the valleys separately. In each of the them we may introduce another time-reversion operation: $H\to \sigma_y H^T \sigma_y$. This time-reversal symmetry is broken by any mass; hence, the mass term acts as an effective magnetic field in the single-valley Dirac Hamiltonian \cite{Semenov84, Volkov85, Pankratov87, Haldane88}.  

It is well known that the single-valley Hall conductivity inside the gap reads $\sigma_{xy} = (2e^2/h) \sign(m)/2$ \cite{Haldane88,Ludwig94,ostrovsky07,evers08}. The integer quantum Hall phase transition at zero mass corresponds to a jump in $\sigma_{xy}$ of the size of the conductance quantum. Thus, the zero-mass line can be viewed as the boundary between insulators with Chern numbers different by 1 and has to contain exactly one topologically protected chiral channel \cite{Volkov85,VolovikBook2003,Qi2011}.  The chiral electron propagates along the boundary in one direction in one valley but in the opposite direction in the other valley independent of the potentials $V$ and $\bb{A}$ \cite{Tudor12}. Thus, the conductance of such a chiral channel at the Dirac point equals $2e^2/h$, where the factor $2$ is due to the spin degeneracy. 

The topological protection of the chiral mode can, however, be destroyed either by inter-valley scattering or by tunnelling into a neighbouring zero-mass line in the same valley, which propagates back into incoming lead. The former processes are negligible while the later ones are suppressed in the limit $\xi \Delta/\hbar v \gg 1$ (provided $\sqrt{\la V^2\ra} \ll \Delta$).  It is worth noting that chiral states with zero energy are absent in closed loops (vortices) because the periodic boundary condition with the Berry phase $\pi$ cannot be fulfilled for the chiral states due to the absence of a dynamic phase \cite{Bardarson10}. 

\begin{figure}[ht]
\centerline{\includegraphics[width=0.9\columnwidth]{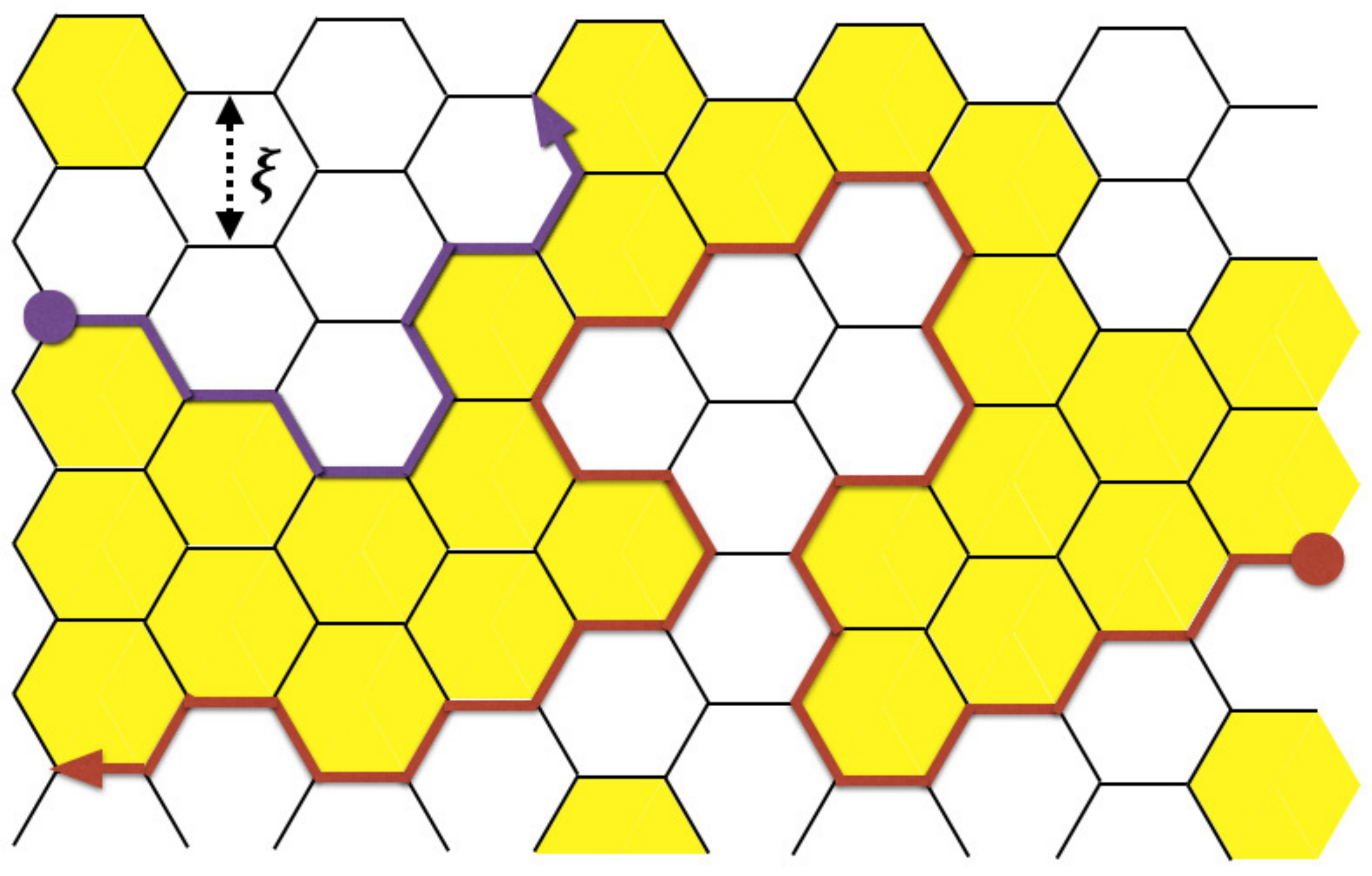}}
\caption{Percolation model. Hexagons are coloured yellow with a probability $p$, and the boundary between white and yellow hexagons corresponds to a zero-mass line that supports two counter-propagating chiral edge states in different valleys. Each line connecting the leads contributes a conductance quantum of $2e^2/h$. The statistics of the zero-mass lines is described by the Schramm-Loewner evolution \cite{Schramm}.}
%\vspace*{-0.5cm}
\label{fig:walk}
\end{figure}

Let us now propose a simple percolation model to study transport at charge neutrality in a correlated mass landscape.  We consider bond percolation on an effective honeycomb lattice. (The symmetry of the lattice is irrelevant since we consider a universal continuum limit of a large system size.)  Cells of an effective honeycomb lattice with a period $\xi$ are coloured yellow with a probability $p$ as illustrated in Fig.\,\ref{fig:walk}. The boundary between yellow and white regions corresponds to a line of zero mass. Each line connecting the leads contributes the conductance quantum $2e^2/h$. Thus, the mean conductivity can be inferred from the statistical analysis of boundaries of the classical percolation cluster in Fig.~\ref{fig:walk} as
\be
\label{sigma}
\sigma = \frac{2e^2}{h}\,\frac{L \la N_\textrm{line} \ra}{W},
\e
where $W$ and $L$ are the width and the length of a rectangular graphene sample, respectively, and $\la N_\textrm{line} \ra$ stands for the the average number of zero-mode lines connecting sample edges at $x=0$ and $x=L$ (the leads). The conductivity (\ref{sigma}) must approach the actual 2D conductivity of the system in the limit $W\gg L$. 

%%%%%%%%%%%%%%%%%%%%%%%%%%%%
%%%% fig:walk
%%%%%%%%%%%%%%%%%%%%%%%%%%%%
%%%%%%%%%%%%%%%%%%%%%%%%%%%%% 

The percolation problem for a lattice of randomly coloured hexagons is very well studied. It is well known that the probability of having an infinite cluster of the same colour hexagons is vanishing unless $p=1/2$ \cite{Cardy}, which is qualitatively identical to the behaviour of the quantum model in the class $A$. The properties of critical percolation at $p=1/2$, which corresponds to $\la m\ra=0$, are universal in the thermodynamic limit, i.e., do not depend on the underlying lattice model. 

A number of exact results for the critical percolation have been obtained by means of the conformal field theory \cite{Cardy,Saleur87,Isichenko92}. More recently the stochastic approach known as the Schramm-Loewner evolution (SLE) has been developed as a mathematically rigorous alternative to tackle the problem \cite{Schramm}. In this formulation a zero-mass line at the cluster boundary is viewed as a random walk. The statistical properties of such a walk at $p=1/2$ yield the stochastic evolution in the class SLE$_6$ \cite{Smirnov01,Beffara04,Kager04}. For instance, it can be rigorously proven that the line connecting the leads has a fractal dimension $7/4$, instead of $2$ as for Brownian motion, which means $L_\textrm{line} \propto L (L/\xi)^{3/4}$ in the limit $L\gg \xi$. 

\begin{figure}[ht]
\centerline{\includegraphics[width=0.95\columnwidth]{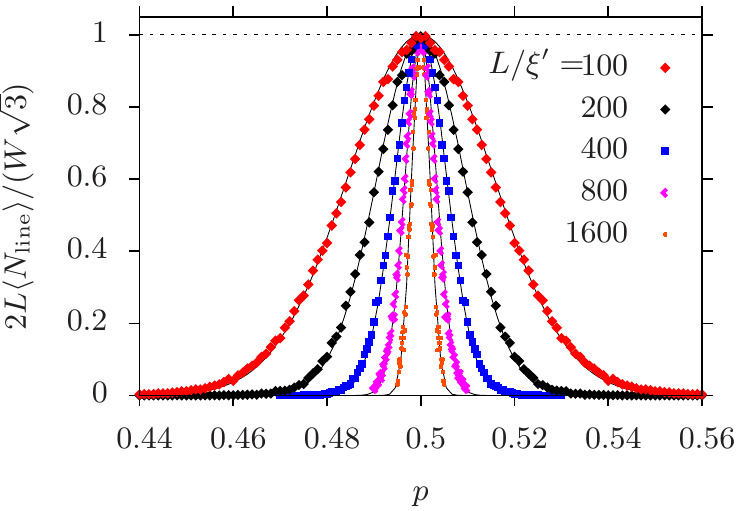}}
\caption{The average number of zero-mode lines connecting the leads at $x=0$ and $x=L$ in the random walk model on a hexagonal lattice (see Fig.~\ref{fig:walk}) with periodic boundary conditions in $y$. The dots represent numerical data obtained for $W/L=2$. Solid lines correspond to Eq.~(\ref{num}).}
%\vspace*{-0.5cm}
\label{fig:P}
\end{figure}

One of the seminal Cardy formulas \cite{Cardy} provides a universal expression for the mean number of independent percolation clusters $\la N_c\ra$ connecting two arbitrary arcs (see also Ref.~\cite{Smirnov11}). This number is determined by a single parameter which encodes the shape of the arcs. For rectangular geometry in the limit $W\gg L$ the number of percolating lines is 2 times larger than the number of clusters in the leading order in $W/L$. In this case the Cardy formula gives $\la N_\textrm{line} \ra = 2   \la N_c\ra = \sqrt{3} W/2 L$. We illustrate this result with straightforward numerical simulations in Fig.~\ref{fig:P} for a sample with periodic boundary conditions in the $y$ direction. 

We also find that the distribution of the line lengths is log-normal and
\be
\label{log}
\exp \lt\la \ln (L_\textrm{line}/L)\rt\ra = c (L/\xi')^{3/4},\qquad \xi'=\sqrt{3}\xi/2,
\e
where  $c\approx 0.7$.  Away from $p=1/2$, the average number of lines between the leads for $W\gg L$ is very well fitted by the Gaussian law
\be
\label{num}
\la N_\textrm{line} \ra = \frac{\sqrt{3} W}{2 L}\exp\lt[-\frac{(p-1/2)^2}{p_0^2}\rt], \quad 
\e
where $p_0 = c(\xi'/L)^{3/4}$ with $c \approx 0.7$ (see Fig.~\ref{fig:P}). 

%%%%%%%%%%%%%%%%%%%%%%%%%%%%
%%%% fig:walk
%%%%%%%%%%%%%%%%%%%%%%%%%%%%
%%%%%%%%%%%%%%%%%%%%%%%%%%%%% 

Critical percolation for $p=1/2$ corresponds to a system with $\la m \ra=0$; moreover, $p-1/2 \propto\la m \ra /\Delta$.  The insulating behaviour observed in Ref.~\onlinecite{Woods14} can be explained by the finite average mass, i.e. $p\neq 1/2$. It might be difficult in an experiment to ensure the conditions required for the classical percolation regime, which explains the scarce number of samples that demonstrate insulating behaviour on small scales \cite{Woods14}.  

A slight increase in orientation angle $\phi \sim 1\degree$ makes the mean mass negligible, while the amplitude $\Delta$ remains large. In this regime we can expect a critical state with $p=1/2$, which is characterised by the conductivity $\sigma^\textrm{perc}_{xx} = \sqrt{3} e^2/h$. This value lies in between the conductivity fixed points for the fully quantum models: $\sigma^D_{xx}$ and $\sigma^\textrm{QHE}_{xx}$. Fully misaligned samples correspond to $m \ll V$. In this case the system at $\ep=0$ has a tendency to flow towards the symplectic metal due to the leading role of the scalar potential $V$  \cite{ostrovsky06,titov07}. 

The results of Eqs.~(\ref{log}) and (\ref{num}) are equivalent to those obtained in the classical percolation model of the quantum Hall transition \cite{Milnikov88,Gurarie01} and can be expected for a critical 2D percolation on any lattice \cite{Cardy,Saleur87,Isichenko92,Smirnov01,Beffara04,Kager04}. It is understood that quantum tunnelling always plays an important role at large scales where a full quantum consideration is necessary. In our case such tunnelling is facilitated by a finite $V$ and $\bb{A}$ potentials, which drive our system in the thermodynamic limit from the classical percolation regime [Eqs.~(\ref{sigma}) and (\ref{num})] to the quantum Hall critical point. This limit may not be of practical importance though, since the size of samples in commensurate phase is technologically restricted to a few microns. The conditions are hard to fulfil in an experiment which explains why it is so difficult to observe insulating behaviour in finite samples \cite{Woods14}. 

In conclusion, we present a scenario for the metal-insulator transition in graphene on hBN, which accompanies the commensurate-incommensurate structural transition observed in recent experiments \cite{Woods14}. We find that the charge transport at charge neutrally is governed by a classical percolation model, at least, on intermediate length scales.  Small lattice misalignment corresponds to a critical-metal state with the conductivity $\sigma \approx \sqrt{3}e^2/h$. The best alignment between graphene and hBN lattices leads to the formation of the mean non-zero mass in the corresponding Dirac equation, which is responsible for the insulating behaviour.  
 
We are grateful to Carlo Beenakker, Sergey Brener, Andre Geim, Igor Gornyi, Alexander Mirlin, Kostya Novoselov, and Stanislav Smirnov for helpful discussions and to Paul Kelly for bringing our attention to the results of Ref.~\cite{Bokdam13} before publication. The work was supported by the Dutch Science Foundation NWO/FOM 13PR3118, the ERC Advanced Grant No. 338957 FEMTO/NANO, EU FP7 Graphene Flagship Grant No. 604391 and by the EU Network Grant InterNoM.

\end{document}